\documentclass[sigconf,nonacm]{acmart}

\renewcommand\footnotetextcopyrightpermission[1]{}
\usepackage{booktabs}
\usepackage{multirow}
\usepackage{url}
\usepackage{amsmath}
\usepackage{graphicx}
\usepackage{subcaption}
\usepackage{tikz}
\usetikzlibrary{arrows.meta,positioning,fit,backgrounds,calc,decorations.pathreplacing,shapes.geometric}

\begin{document}

\title{MEMOIR: Temporal Behavioral Memory for Recommendation Across the Preference-Drift Spectrum}

\author{Younggue Bae\footnotemark[1]}
\affiliation{%
  \institution{Independent Researcher}
  \city{Vancouver}
  \country{Canada}
}
\email{louiezzang@gmail.com}

\begin{abstract}
We propose MEMOIR, a framework that segments user interaction histories into temporal windows, generates semantic behavioral memory for each period using an LLM, and aggregates current state, evolution direction, and predicted future into a single user representation.
On the Electronics and Clothing\_Shoes\_and\_Jewelry categories of Amazon Reviews 2023, MEMOIR is statistically tied with UniSRec, the strongest baseline, on aggregate NDCG@10 (0.0643 vs.\ 0.0641), splitting the four reported metrics 2--2: MEMOIR leads NDCG@10 and MRR, UniSRec leads HR@10 and HR@20.
An ablation study finds that no single architectural component---the evolution-preserving contrastive loss, its directional-consistency term, or temporal window segmentation itself---individually explains much of MEMOIR's $\approx$18\% relative gain over ID-based SASRec; all four ablations land within 2\% of the full model on aggregate NDCG@10.
Stratifying test performance by a composite preference-drift score instead reveals where the gain concentrates: MEMOIR leads on ranking-quality metrics (NDCG@10, MRR) specifically among users at the high- and low-drift extremes of the distribution, while UniSRec leads the volume-oriented HR@10/HR@20 metrics across all drift strata and edges out MEMOIR on ranking quality in the middle band.
We report this drift-stratified pattern, rather than the near-tied aggregate numbers or any single ablated component, as MEMOIR's most substantive and reproducible finding, and surface why it holds as an open question for future work.
\end{abstract}



\maketitle
\pagestyle{plain}
\footnotetext[1]{Code: \url{https://github.com/louiezzang/MEMOIR}}

\section{Introduction}

\begin{sloppypar}
Sequential recommendation models such as SASRec~\cite{kang2018sasrec} and BERT4Rec~\cite{sun2019bert4rec} capture user preferences from interaction histories, implicitly learning behavioral patterns through item-level sequence processing.
Recent LLM-enhanced methods like SRA-CL~\cite{cui2025sracl} further enrich representations with semantic knowledge from large language models.
However, these approaches treat user preference as either a static summary or a byproduct of next-item prediction---none explicitly models \emph{how} preferences evolve over time.
\end{sloppypar}

We argue that preference evolution deserves first-class modeling.
Two users who both currently purchase athletic wear may have very different recommendation needs if one is a long-time fitness enthusiast while the other just started exercising.
A static representation or a single LLM-generated summary cannot distinguish these trajectories.
Empirically, we find that this drift is pervasive: on the Electronics and Clothing\_Shoes\_and\_Jewelry categories of Amazon Reviews 2023, the mean Jensen-Shannon divergence between a user's adjacent monthly category distributions is 0.795 (out of 1.0), and 92.2\% of items interacted with in any given window are items the user has never engaged with before.
Even the bottom quartile of users by drift score exhibits substantial preference change, suggesting that evolution-aware modeling benefits nearly all users, not only an extreme minority.

We propose \textbf{MEMOIR} (\textbf{M}emory-based \textbf{E}volution \textbf{MO}del for \textbf{I}nference in \textbf{R}ecommendation), a framework that:
\begin{enumerate}
    \item Segments user interaction histories into temporal windows and generates semantic behavioral memory for each period using an LLM encoder;
    \item Learns the evolution trajectory through a novel \emph{evolution-preserving contrastive loss} that enforces both temporal smoothness and directional consistency;
    \item Aggregates user representations by fusing current state, evolution direction, and predicted future into an evolution-aware memory.
\end{enumerate}
We test each of these three components individually via ablation (§\ref{sec:ablation}); as we report there, none in isolation explains most of MEMOIR's aggregate gain over ID-based baselines, and the more robust, reproducible effect we identify is concentrated in a specific part of the user population rather than uniform across it (§\ref{sec:drift_analysis}).

More fundamentally, existing LLM-enhanced methods remain ID-based sequential models at their core: the LLM generates offline summaries or augmented views that improve training quality, but the recommendation backbone still operates on item-ID embeddings, and any contrastive or semantic signal serves as an auxiliary that is discarded at inference time.
MEMOIR inverts this relationship: the LLM \emph{is} the user encoder, contrastive learning is the mechanism through which the structure of preference evolution is learned, and the evolution-aware representation persists through to the final recommendation score.
Our key insight is that \emph{recommendation should target a user's evolving trajectory, not just their current state}---and that trajectory modeling requires the LLM to be a first-class architectural component, not an offline preprocessing step.

\section{Related Work}

\paragraph{Sequential and Temporal User Modeling.}
Early sequential recommendation models capture behavioral patterns via recurrent networks~\cite{hidasi2016gru4rec} or self-attention~\cite{kang2018sasrec,sun2019bert4rec}, producing a single implicit representation of the user's full history without explicitly modeling change over time.
Time-aware extensions such as TiSASRec~\cite{li2020tisasrec} incorporate explicit time-interval embeddings between consecutive interactions, enabling sensitivity to recency and interaction gaps, but still collapse history into a single aggregated state.
SURGE~\cite{chang2021surge} constructs dynamic interest graphs that cluster items into soft interest categories, providing coarse-grained temporal grouping of evolving interests.
Unlike these methods, MEMOIR treats the \emph{trajectory} of change as the primary modeling target: temporal windows yield a sequence of semantic memory states, and contrastive losses are applied directly to the directions of change between those states.

\paragraph{LLM-Based Recommendation.}
Large language models have been applied to recommendation as frozen feature encoders~\cite{cui2025sracl} and instruction-tuned rankers~\cite{bao2023tallrec,liao2024llara}.
In all of these settings the LLM serves an auxiliary role---enriching an underlying ID-based sequential model---and produces a single, session-level representation with no mechanism to model how preferences evolve across time.
Crucially, the contrastive and semantic objectives in these methods function as training-time tools to improve the ID-based backbone's representation quality; they are discarded at inference.
MEMOIR instead uses the LLM as a \emph{first-class temporal memory encoder}, generating a distinct semantic embedding for each window of activity.
Contrastive learning is not an auxiliary training signal but the mechanism through which the structure of preference evolution---temporal smoothness and directional consistency---is encoded into the representation space and carried through to inference.

\paragraph{Memory-Augmented User Modeling.}
External memory mechanisms~\cite{sukhbaatar2015memnn} have been applied to user modeling to retain long-horizon behavioral context beyond what recurrent hidden states can store.
Memory-augmented recommendation systems~\cite{ebesu2018cmn} maintain differentiable key-value stores that are updated as new interactions arrive, enabling persistent preference tracking.
MEMOIR shares the intuition of maintaining explicit behavioral memory, but departs from slot-based write-and-read architectures: it uses an LLM to distill each temporal window into a semantic summary and then learns the \emph{evolution} of those summaries through contrastive trajectory objectives rather than slot-update rules.

\section{MEMOIR Framework}

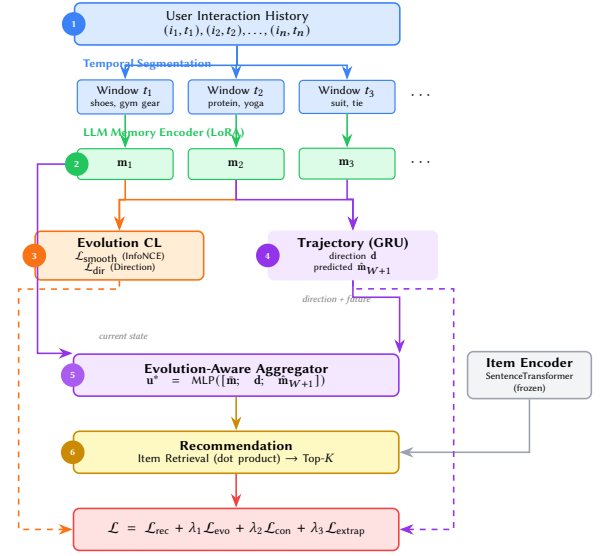
\begin{figure}[t]
    \centering
    \resizebox{\columnwidth}{!}{%
\definecolor{inputblue}{RGB}{219,234,254}
\definecolor{inputborder}{RGB}{59,130,246}
\definecolor{encodergreen}{RGB}{220,252,231}
\definecolor{encoderborder}{RGB}{34,197,94}
\definecolor{contrastorange}{RGB}{255,237,213}
\definecolor{contrastborder}{RGB}{249,115,22}
\definecolor{aggpurple}{RGB}{243,232,255}
\definecolor{aggborder}{RGB}{147,51,234}
\definecolor{recyellow}{RGB}{254,249,195}
\definecolor{recborder}{RGB}{202,138,4}
\definecolor{lossred}{RGB}{254,226,226}
\definecolor{lossborder}{RGB}{239,68,68}
\definecolor{graybox}{RGB}{243,244,246}
\definecolor{grayborder}{RGB}{156,163,175}

\begin{tikzpicture}[
    node distance=0.55cm and 0.6cm,
    >={Stealth[length=2.5mm]},
    font=\sffamily\small,
    box/.style={
        rectangle, rounded corners=3pt,
        draw=#1, fill=#1!15, line width=0.8pt,
        minimum height=0.75cm, minimum width=2.6cm,
        text width=2.4cm, align=center,
        font=\sffamily\footnotesize
    },
    widebox/.style={
        rectangle, rounded corners=3pt,
        draw=#1, fill=#1!15, line width=0.8pt,
        minimum height=0.75cm, minimum width=5.8cm,
        text width=5.6cm, align=center,
        font=\sffamily\footnotesize
    },
    smallbox/.style={
        rectangle, rounded corners=2pt,
        draw=#1, fill=#1!10, line width=0.6pt,
        minimum height=0.55cm, minimum width=1.6cm,
        text width=1.5cm, align=center,
        font=\sffamily\scriptsize
    },
    arrow/.style={->, thick, color=#1},
    darrow/.style={->, thick, dashed, color=#1},
    label/.style={font=\sffamily\scriptsize\bfseries, color=#1},
]

\node[widebox=inputborder, fill=inputblue] (input)
    {User Interaction History\\[-1pt]
    {\scriptsize $(i_1,t_1), (i_2,t_2), \ldots, (i_n,t_n)$}};

\node[below=0.5cm of input] (seglab) {};

\node[smallbox=inputborder, fill=inputblue, below=0.6cm of input, xshift=-2.0cm] (w1)
    {Window $t_1$\\[-1pt] {\tiny shoes, gym gear}};
\node[smallbox=inputborder, fill=inputblue, right=0.25cm of w1] (w2)
    {Window $t_2$\\[-1pt] {\tiny protein, yoga}};
\node[smallbox=inputborder, fill=inputblue, right=0.25cm of w2] (w3)
    {Window $t_3$\\[-1pt] {\tiny suit, tie}};
\node[right=0.15cm of w3, font=\sffamily\small] (wdots) {$\cdots$};

\node[label=inputborder, above=0.05cm of w1.north west, anchor=south west] {Temporal Segmentation};

\draw[arrow=inputborder] (input.south) -- ++(0,-0.35) -| (w1.north);
\draw[arrow=inputborder] (input.south) -- ++(0,-0.35) -| (w2.north);
\draw[arrow=inputborder] (input.south) -- ++(0,-0.35) -| (w3.north);

\node[smallbox=encoderborder, fill=encodergreen, below=0.6cm of w1] (m1)
    {$\mathbf{m}_1$};
\node[smallbox=encoderborder, fill=encodergreen, below=0.6cm of w2] (m2)
    {$\mathbf{m}_2$};
\node[smallbox=encoderborder, fill=encodergreen, below=0.6cm of w3] (m3)
    {$\mathbf{m}_3$};
\node[right=0.15cm of m3, font=\sffamily\small] (mdots) {$\cdots$};

\node[label=encoderborder, above=0.05cm of m1.north west, anchor=south west]
    {LLM Memory Encoder (LoRA)};

\draw[arrow=encoderborder] (w1) -- (m1);
\draw[arrow=encoderborder] (w2) -- (m2);
\draw[arrow=encoderborder] (w3) -- (m3);


\node[box=contrastborder, fill=contrastorange, below left=0.9cm and -0.3cm of m2,
      text width=2.8cm, minimum width=3.0cm] (evo)
    {\textbf{Evolution CL}\\[-2pt]
    {\tiny $\mathcal{L}_\text{smooth}$ (InfoNCE)}\\[-2pt]
    {\tiny $\mathcal{L}_\text{dir}$ (Direction)}};

\node[box=aggpurple, fill=aggpurple, below right=0.9cm and -0.3cm of m2,
      text width=2.8cm, minimum width=3.0cm] (traj)
    {\textbf{Trajectory (GRU)}\\[-2pt]
    {\tiny direction $\mathbf{d}$}\\[-2pt]
    {\tiny predicted $\hat{\mathbf{m}}_{W+1}$}};

\draw[arrow=contrastborder] (m1.south) -- ++(0,-0.35) -| (evo.north);
\draw[arrow=contrastborder] (m2.south) -- ++(0,-0.35) -| (evo.north);
\draw[arrow=aggborder] (m2.south) -- ++(0,-0.35) -| (traj.north);
\draw[arrow=aggborder] (m3.south) -- ++(0,-0.35) -| (traj.north);

\node[widebox=aggborder, fill=aggpurple,
      below=1.3cm of $(evo.south)!0.5!(traj.south)$] (agg)
    {\textbf{Evolution-Aware Aggregator}\\[-2pt]
    {\scriptsize $\mathbf{u}^* = \text{MLP}([\bar{\mathbf{m}};\; \mathbf{d};\; \hat{\mathbf{m}}_{W+1}])$}};

\draw[arrow=aggborder] (traj.south) -- ++(0,-0.3) -| (agg.north east);
\draw[arrow=aggborder] (m1.west) -- ++(-0.7,0) |- (agg.north west);

\node[font=\sffamily\tiny\itshape, color=gray, below=0.15cm of traj.south, xshift=-0.3cm] {direction + future};
\node[font=\sffamily\tiny\itshape, color=gray] at ($(agg.north west)+(0.9,0.3)$) {current state};

\node[widebox=recborder, fill=recyellow, below=0.6cm of agg] (rec)
    {\textbf{Recommendation}\\[-2pt]
    {\scriptsize Item Retrieval (dot product) $\rightarrow$ Top-$K$}};

\draw[arrow=recborder] (agg) -- (rec);

\node[box=grayborder, fill=graybox, right=1.2cm of agg,
      text width=2.0cm, minimum width=2.2cm] (item)
    {\textbf{Item Encoder}\\[-2pt]
    {\tiny SentenceTransformer}\\[-2pt]
    {\tiny (frozen)}};

\draw[arrow=grayborder] (item.south) |- (rec.east);

\node[widebox=lossborder, fill=lossred, below=0.6cm of rec] (loss)
    {$\mathcal{L} = \mathcal{L}_\text{rec} + \lambda_1 \mathcal{L}_\text{evo} + \lambda_2 \mathcal{L}_\text{con} + \lambda_3 \mathcal{L}_\text{extrap}$};

\draw[arrow=lossborder] (rec) -- (loss);
\draw[darrow=contrastborder] (evo.south) -- ++(0,-0.15) -- ++(-1.8,0) |- (loss.west);
\draw[darrow=aggborder] (traj.south) -- ++(0,-0.15) -- ++(1.8,0) |- (loss.east);

\node[font=\sffamily\tiny\bfseries, circle, fill=inputborder, text=white,
      inner sep=1pt, minimum size=12pt] at (input.west) {\textbf{1}};
\node[font=\sffamily\tiny\bfseries, circle, fill=encoderborder, text=white,
      inner sep=1pt, minimum size=12pt] at (m1.west) {\textbf{2}};
\node[font=\sffamily\tiny\bfseries, circle, fill=contrastborder, text=white,
      inner sep=1pt, minimum size=12pt] at (evo.west) {\textbf{3}};
\node[font=\sffamily\tiny\bfseries, circle, fill=aggborder, text=white,
      inner sep=1pt, minimum size=12pt] at (traj.west) {\textbf{4}};
\node[font=\sffamily\tiny\bfseries, circle, fill=aggborder!80, text=white,
      inner sep=1pt, minimum size=12pt] at (agg.west) {\textbf{5}};
\node[font=\sffamily\tiny\bfseries, circle, fill=recborder, text=white,
      inner sep=1pt, minimum size=12pt] at (rec.west) {\textbf{6}};

\end{tikzpicture}%
    }
    \caption{MEMOIR architecture: user history is segmented into temporal windows, encoded into behavioral memory via an LLM, and learned through evolution-preserving contrastive loss. The evolution-aware aggregator fuses current state, direction, and predicted future.}
    \label{fig:architecture}
\end{figure}

Figure~\ref{fig:architecture} illustrates the overall architecture.
Given a user's time-stamped interaction history, MEMOIR processes it through four stages: temporal segmentation, memory encoding, evolution-preserving contrastive learning, and evolution-aware aggregation.

\subsection{Temporal Behavioral Memory}

Given a user $u$'s interaction history $\mathcal{H}_u = \{(i_1, t_1), \ldots, (i_n, t_n)\}$, we partition it into $W$ temporal windows based on calendar months.
Each window $w_t$ is serialized into a natural language description (e.g., \emph{``User highly rated `running shoes' in Sports; purchased `protein powder' in Health''}).
An LLM encoder $f_\theta$ with LoRA fine-tuning maps each window text to a semantic memory embedding:
\begin{equation}
    \mathbf{m}_t = \text{Proj}(\text{MeanPool}(f_\theta(w_t))) \in \mathbb{R}^D
\end{equation}
where $\text{Proj}$ is a two-layer MLP projecting the LLM hidden states to the embedding space.
The result is a memory sequence $\mathbf{M}_u = [\mathbf{m}_1, \ldots, \mathbf{m}_W]$ representing the user's behavioral evolution.

\subsection{Evolution-Preserving Contrastive Loss}

We introduce a contrastive loss $\mathcal{L}_\text{evo}$ comprising two complementary terms:

\paragraph{Temporal Smoothness.}
Adjacent windows from the same user form positive pairs, enforcing continuity:
\begin{equation}
    \mathcal{L}_\text{smooth} = -\log \frac{\exp(\text{sim}(\mathbf{m}_t, \mathbf{m}_{t+1}) / \tau)}{\sum_{j \in \mathcal{N}} \exp(\text{sim}(\mathbf{m}_t, \mathbf{m}_j) / \tau)}
\end{equation}
where $\mathcal{N}$ includes windows from other users as negatives and $\tau$ is the temperature.

\paragraph{Directional Consistency.}
Beyond smoothness, we enforce that the \emph{direction} of evolution remains coherent:
\begin{equation}
    \mathcal{L}_\text{dir} = \text{ReLU}\left(\gamma - \cos(\mathbf{d}_{t-1}, \mathbf{d}_t)\right)
\end{equation}
where $\mathbf{d}_t = \hat{\mathbf{m}}_{t+1} - \hat{\mathbf{m}}_t$ is the evolution direction vector computed on $\ell_2$-normalized memories $\hat{\mathbf{m}} = \mathbf{m}/\|\mathbf{m}\|_2$, and $\gamma$ is a margin hyperparameter.
This loss encodes the inductive bias that if a user's preferences shift from casual wear to sportswear, consecutive direction vectors should be aligned.

The combined evolution loss is:
\begin{equation}
    \mathcal{L}_\text{evo} = \mathcal{L}_\text{smooth} + \alpha \cdot \mathcal{L}_\text{dir}
\end{equation}

\subsection{Trajectory Prediction}

A GRU processes the window memory sequence and emits a final hidden state:
\begin{equation}
    \mathbf{h}_W = \text{GRU}([\mathbf{m}_1, \ldots, \mathbf{m}_W])
\end{equation}
A two-layer MLP projection head maps this hidden state to a predicted next-step memory:
\begin{equation}
    \hat{\mathbf{m}}_{W+1} = \text{MLP}_\text{pred}(\mathbf{h}_W)
\end{equation}
The evolution direction is computed directly from consecutive window embeddings:
\begin{equation}
    \mathbf{d} = \mathbf{m}_W - \mathbf{m}_{W-1}
\end{equation}

We supervise the predictor with a trajectory extrapolation loss using hold-out-last-window supervision: during training the GRU is applied to the first $W-1$ windows to predict $\hat{\mathbf{m}}_W$, which is then compared to the held-out actual embedding:
\begin{equation}
    \mathcal{L}_\text{extrap} = \text{InfoNCE}(\hat{\mathbf{m}}_W, \mathbf{m}_W)
\end{equation}
The positive for each example is the held-out window $\mathbf{m}_W$ of the same user; negatives are the held-out windows of all other users in the same mini-batch (in-batch negatives, following the standard InfoNCE construction~\cite{oord2018cpc}).

\subsection{Evolution-Aware Aggregation}

Instead of a simple weighted mean of window memories, we fuse three signals via an MLP:
\begin{equation}
    \mathbf{u}^* = \text{MLP}([\bar{\mathbf{m}}; \mathbf{d}; \hat{\mathbf{m}}_{W+1}])
\end{equation}
where $\bar{\mathbf{m}}$ is the recency-weighted aggregate (current state), $\mathbf{d}$ is the direction vector (where from), and $\hat{\mathbf{m}}_{W+1}$ is the predicted future (where to).
This ensures that users with identical current preferences but different evolution paths receive distinct representations.

\subsection{Behavioral Consistency Loss}

To anchor the evolution-aware user representation to the item embedding space, we add a consistency regularizer that penalises the squared distance between the normalized aggregate memory and the normalized target item embedding:
\begin{equation}
    \mathcal{L}_\text{con} = \left\|
        \frac{\mathbf{u}^*}{\|\mathbf{u}^*\|_2}
        -
        \frac{\mathbf{e}_i}{\|\mathbf{e}_i\|_2}
    \right\|_2^2
\end{equation}
where $\mathbf{e}_i$ is the target item embedding, treated as a fixed target (gradients are not propagated through it in this term).
This loss prevents the evolution trajectory from drifting into regions of the embedding space that are semantically detached from the items the user actually interacts with.

\subsection{Training Objective}

The full training loss combines recommendation, evolution, consistency, and extrapolation:
\begin{equation}
    \mathcal{L} = \mathcal{L}_\text{rec} + \lambda_1 \mathcal{L}_\text{evo} + \lambda_2 \mathcal{L}_\text{con} + \lambda_3 \mathcal{L}_\text{extrap}
\end{equation}
where $\mathcal{L}_\text{rec}$ is an InfoNCE recommendation loss over user-item dot products.

\subsection{Design Assumptions and Limitations}
\label{sec:limitations}

\paragraph{Directional consistency assumption.}
$\mathcal{L}_\text{dir}$ encodes the inductive bias that consecutive evolution direction vectors should be aligned, which suits users with monotonic drift (e.g., a gradual shift from casual to formal wear).
This assumption is violated for users with oscillating or seasonal preferences (e.g., alternating between summer and winter categories), where consecutive direction vectors are naturally opposed.
To mitigate this, $\mathcal{L}_\text{dir}$ fires only when at least three consecutive windows are available and uses a hinge formulation with margin $\gamma$, so a small degree of direction reversal is tolerated without penalty.
Users with strongly oscillating preferences may nonetheless be hurt by this term; reducing $\alpha$ or disabling $\mathcal{L}_\text{dir}$ is advisable in such domains.

\paragraph{Sparse-user degradation.}
Monthly calendar segmentation yields fewer windows for low-activity users.
MEMOIR degrades gracefully: for users with a single window, the direction vector $\mathbf{d}$ is set to zero (no evolution signal), and $\mathcal{L}_\text{dir}$ and $\mathcal{L}_\text{extrap}$ are not computed (they require $\geq 2$ and $\geq 3$ windows respectively).
In these cases the model reduces to an LLM-encoded single-state representation with the recommendation loss $\mathcal{L}_\text{rec}$ and consistency loss $\mathcal{L}_\text{con}$ as the sole training signal, which remains competitive with standard LLM-based baselines.

\subsection{Complexity Analysis}

\paragraph{Training.}
The dominant cost is the LLM memory encoder: encoding all user windows requires $\mathcal{O}(|\mathcal{U}| \cdot W \cdot L \cdot d_\text{LLM})$ FLOPs, where $|\mathcal{U}|$ is the number of users, $W$ is the maximum number of windows, $L$ is average tokens per window, and $d_\text{LLM}$ is the LLM hidden dimension.
With \texttt{freeze\_llm=True} this cost is paid once and fully cached before training begins; subsequent training steps incur only $\mathcal{O}(W \cdot h^2)$ for the GRU (hidden size $h$) and $\mathcal{O}(B \cdot K)$ for the contrastive losses (batch size $B$, negatives per anchor $K$).
With LoRA fine-tuning, the LLM forward pass repeats each epoch but operates only over trainable adapter parameters ($r \ll d_\text{LLM}$), keeping the added parameter count at $\mathcal{O}(r \cdot d_\text{LLM})$ per adapted layer.

\paragraph{Inference.}
Given cached or freshly computed window embeddings, inference requires a single GRU pass ($\mathcal{O}(W \cdot h)$) and one MLP aggregation step ($\mathcal{O}(3D \cdot d_\text{agg})$), yielding a user representation in $\mathcal{O}(W \cdot D)$ time---linear in the number of windows and independent of raw interaction count.
Storage is $\mathcal{O}(|\mathcal{U}| \cdot W \cdot D)$ for the memory grid, comparable to standard matrix-factorization user tables.

\section{Experiments}

\subsection{Setup}

\paragraph{Dataset.}
We use the Electronics and Clothing\_Shoes\_and\_Jewelry categories of Amazon Reviews 2023~\cite{hou2024bridging}, filtering users with $\geq$5 interactions.
After temporal segmentation with monthly windows, the dataset contains 99,355 users with 569,397 behavioral windows (mean 5.7 windows per user).
We use an 80/10/10 chronological train/validation/test split.
A pre-training drift analysis (§\ref{sec:drift_analysis}) classifies users into high (top 25\%), medium (middle 50\%), and low (bottom 25\%) drift groups based on a composite score of category JSD, rating variance, and novelty rate; mean composite drift score is 0.695.

\paragraph{Baselines.}
We compare against seven baselines spanning four categories:
(1)~\emph{Traditional sequential}: GRU4Rec~\cite{hidasi2016gru4rec}, SASRec~\cite{kang2018sasrec};
(2)~\emph{Contrastive sequential}: CL4SRec~\cite{xie2022cl4srec}, DuoRec~\cite{qiu2022duorec};
(3)~\emph{Text-aware sequential}: SASRec-Text (SASRec with MiniLM text embeddings replacing learned ID embeddings), UniSRec~\cite{hou2022unisrec} (whitening transform and mixture-of-experts adaptor over pretrained text representations);
(4)~\emph{LLM-enhanced contrastive}: SRA-CL~\cite{cui2025sracl}.
All baselines are trained with a full-catalog softmax cross-entropy recommendation objective, scoring each user representation against the entire item catalog rather than a sampled subset; this standardizes the training signal across methods so that reported differences reflect architectural choices rather than negative-sampling scheme, and departs from some baselines' original training procedures (e.g., SASRec and GRU4Rec originally used sampled or in-batch negative objectives).
The text-aware baselines use the same pretrained encoder (all-MiniLM-L6-v2) as MEMOIR's item encoder, ensuring that any performance gap reflects architectural differences rather than encoder quality.
Notably, SASRec-Text serves as a controlled probe: it tests whether replacing learned ID embeddings with frozen semantic embeddings alone improves recommendation, without any evolution modeling or alignment training.

\paragraph{Implementation.}
\begin{sloppypar}
MEMOIR uses TinyLlama-1.1B-Chat~\cite{zhang2024tinyllama} as the memory encoder and all-MiniLM-L6-v2 as the frozen item encoder.
Embedding dimension $D$=256, temperature $\tau$=0.07, $\lambda_1$=0.5, $\lambda_2$=0.3, $\lambda_3$=0.3, $\alpha$=0.2.
All experiments use the \emph{frozen LLM} mode: the LLM is held fixed, window embeddings are pre-computed once and cached, and only the projection MLP, GRU, and aggregator are trained end-to-end (AdamW, lr=1e-4, weight decay=0.01).
We train for up to 100 epochs with early stopping (patience=15) based on validation NDCG@10.
\end{sloppypar}

\paragraph{Metrics.}
HR@$K$, NDCG@$K$ ($K \in \{5, 10, 20\}$), and MRR, following the sampled evaluation protocol with 999 negative items.

\subsection{Main Results}

\begin{table}[t]
\caption{Recommendation performance on Amazon Reviews. Best results in \textbf{bold}, second best \underline{underlined}.}
\label{tab:main_results}
\small
\begin{tabular}{lcccc}
\toprule
\textbf{Model} & \textbf{HR@10} & \textbf{NDCG@10} & \textbf{HR@20} & \textbf{MRR} \\
\midrule
GRU4Rec        & 0.0258 & 0.0137 & 0.0395 & 0.0146 \\
SASRec         & 0.0888 & 0.0543 & 0.1198 & 0.0500 \\
CL4SRec        & 0.0892 & 0.0541 & 0.1214 & 0.0497 \\
DuoRec         & 0.0538 & 0.0345 & 0.0729 & 0.0348 \\
\midrule
SASRec-Text    & 0.0810 & 0.0415 & 0.1339 & 0.0415 \\
UniSRec        & \textbf{0.1176} & \underline{0.0641} & \textbf{0.1791} & \underline{0.0603} \\
SRA-CL$^\dagger$ & 0.0340 & 0.0183 & 0.0573 & 0.0201 \\
\midrule
\textbf{MEMOIR (Ours)} & \underline{0.1084} & \textbf{0.0643} & \underline{0.1594} & \textbf{0.0617} \\
\bottomrule
\end{tabular}
{\footnotesize $^\dagger$SRA-CL was originally evaluated on smaller Amazon subcategories (Sports, Beauty, Office) with batch size 256; our setting uses the Electronics and Clothing\_Shoes\_and\_Jewelry categories of Amazon Reviews 2023 (99k users) with batch size 16, reducing effective InfoNCE negatives from 512 to 32.}
\end{table}

Table~\ref{tab:main_results} shows the overall recommendation performance.
MEMOIR and UniSRec split the four metrics 2--2: MEMOIR leads on NDCG@10 and MRR, while UniSRec leads on the wider-cutoff, recall-oriented HR@10 and HR@20; NDCG@10, the metric used for model selection, is effectively a statistical tie (0.0643 vs.\ 0.0641).
Both models clearly outperform the remaining six baselines on every metric.

\paragraph{Semantic features alone do not help.}
A striking finding is that SASRec-Text \emph{underperforms} vanilla SASRec (NDCG@10: 0.0415 vs.\ 0.0543), despite using the same backbone and a pretrained semantic encoder.
This occurs because MiniLM embeddings are optimized for semantic similarity between sentences, which does not directly correspond to behavioral co-occurrence patterns that drive recommendation.
A shallow linear projection is insufficient to bridge this gap: it can rotate the semantic space but cannot align it with the collaborative filtering signal that SASRec's learnable ID embeddings acquire organically through training.
MEMOIR avoids this pitfall through two mechanisms: (1) the behavioral consistency loss $\mathcal{L}_\text{con}$ explicitly anchors the user's semantic representation to the item embedding space, and (2) the evolution-aware aggregation gives the model a richer signal---trajectory, direction, and predicted future---that compensates for any misalignment in the raw semantic features.
SRA-CL, which augments SASRec with LLM-based cross-sequence contrastive learning, similarly underperforms SASRec in our setting (see Table~\ref{tab:main_results}, $^\dagger$), further confirming that LLM augmentation without evolution-aware design does not reliably improve over a well-tuned ID backbone.
Part of this gap is attributable to scale: fitting the full 99k-user set forces a batch size of 16 rather than SRA-CL's original 256, shrinking effective InfoNCE negatives from 512 to 32 ($^\dagger$) and substantially weakening the cross-sequence contrastive signal the method relies on.
But negative count alone does not explain the deficit---the same mechanism identified above for SASRec-Text applies here too: a contrastive term that pulls representations toward LLM-semantic similarity provides no guarantee of alignment with the behavioral co-occurrence signal that drives recommendation, so a weakened version of that signal can interfere with the next-item objective rather than complement it.
This result suggests that the benefit of language model representations in recommendation is not automatic: it requires an architectural mechanism to align semantic and behavioral spaces.

\paragraph{An open question: what explains the remaining gain?}
\begin{sloppypar}
The improvement over the strongest ID-based baseline (SASRec) is substantial ($\approx$18\% relative on NDCG@10), and SASRec-Text underperforming SASRec (Table~\ref{tab:main_results}) rules out pretrained semantics alone as the explanation.
The ablation study below (Table~\ref{tab:ablation}), however, rules out the converse story too: none of MEMOIR's own evolution-modeling components individually explains the gain either.
We do not yet have a component-level account of why MEMOIR outperforms SASRec in aggregate; §\ref{sec:drift_analysis} instead localizes \emph{where} in the user population the gain (and the near-tie with UniSRec) actually comes from.
\end{sloppypar}

\subsection{Ablation Study}
\label{sec:ablation}

\begin{table}[t]
\caption{Ablation study on Amazon Reviews. Best per column in \textbf{bold}, second best \underline{underlined}.}
\label{tab:ablation}
\small
\begin{tabular}{lcc}
\toprule
\textbf{Variant} & \textbf{HR@10} & \textbf{NDCG@10} \\
\midrule
MEMOIR (Full)               & 0.1084 & 0.0643 \\
\quad w/o Evolution CL        & \underline{0.1099} & 0.0642 \\
\quad w/o Direction Loss      & 0.1088 & \underline{0.0644} \\
\quad w/o Temporal Seg.       & 0.1055 & 0.0633 \\
\quad w/ Random-Init Items    & \textbf{0.1121} & \textbf{0.0650} \\
\bottomrule
\end{tabular}
\end{table}

The ``w/ Random-Init Items'' variant replaces the pretrained MiniLM item encoder with randomly initialized ID embeddings, isolating the contribution of MEMOIR's architecture from the pretrained text representations.
Contrary to our expectation going in, Table~\ref{tab:ablation} does not validate any individual component's necessity: all four ablations land within 2\% of the full model's NDCG@10, and \emph{w/ Random-Init Items}---which discards pretrained item semantics entirely---scores best of all five variants on both HR@10 and NDCG@10.
Removing temporal window segmentation (\emph{w/o Temporal Seg.}), the component the model's name and prior framing of this work centered on, costs only 1.6\% relative NDCG@10---an order of magnitude larger than removing the evolution or directional-consistency losses (0.16\% each), but still the same order of magnitude as the 1.1\% gain from \emph{w/ Random-Init Items}.
This mirrors the near-tie with UniSRec in Table~\ref{tab:main_results}: whatever produces MEMOIR's aggregate advantage over SASRec does not isolate cleanly to any one architectural piece we tested here.
We return to this puzzle in the drift-stratified analysis below (§\ref{sec:drift_analysis}), where the gain's location, if not yet its mechanism, becomes clearer.

\subsection{Hyperparameter Sensitivity}

\begin{table}[t]
\caption{Sensitivity of MEMOIR to contrastive temperature $\tau$ and number of windows $W$ (HR@10 on Amazon Reviews). Default values in \textbf{bold}.}
\label{tab:sensitivity}
\small
\begin{tabular}{llc}
\toprule
\textbf{Parameter} & \textbf{Value} & \textbf{HR@10} \\
\midrule
\multirow{3}{*}{$\tau$ (temperature)}
  & 0.03 & 0.1101 \\
  & \textbf{0.07} & \textbf{0.1084} \\
  & 0.2 & 0.1013 \\
\midrule
\multirow{3}{*}{$W$ (max windows)}
  & 3 & 0.1084 \\
  & \textbf{6} & \textbf{0.1084} \\
  & 12 & 0.1084 \\
\bottomrule
\end{tabular}
\end{table}

Table~\ref{tab:sensitivity} reports HR@10 as $\tau$ and $W$ are varied while holding the others at default.
MEMOIR is far more sensitive to the contrastive temperature $\tau$ than to window count $W$, but the pattern is monotonic rather than the U-shape a collapsed-vs.-under-separated-gradient account would predict: HR@10 decreases steadily from $\tau$=0.03 (0.1101, best of the three) through the default $\tau$=0.07 (0.1084) to $\tau$=0.2 (0.1013, an $\approx$8\% relative drop from $\tau$=0.03).
We do not currently have an explanation for why smaller $\tau$ continues to help within this range rather than eventually hurting; retuning the default toward $\tau$=0.03 rather than 0.07 is a low-cost change worth revisiting in future runs.
The number of windows $W$ has no measurable impact in our range (3, 6, and 12 windows are numerically identical to four decimal places), suggesting the aggregator is insensitive to window count once $W \geq 3$, at least on this dataset's activity distribution.
We have not swept the loss weights $\lambda_1$ (evolution) or $\alpha$ (directional consistency) and make no claim about their sensitivity; this is left to future work.

\subsection{Preference Drift Analysis}
\label{sec:drift_analysis}

To characterize the degree of preference evolution in our dataset and validate MEMOIR's design motivation, we compute three per-user drift metrics across temporal windows:
(1)~\emph{Category JSD}: Jensen-Shannon divergence between adjacent windows' category distributions, measuring how much the \emph{types} of items a user engages with shift over time;
(2)~\emph{Rating drift}: variance in mean rating across windows, capturing satisfaction fluctuation;
(3)~\emph{Novelty rate}: fraction of items in each window not seen in any prior window, reflecting exploration breadth.
A composite drift score is derived as the mean of the three min-max-normalized metrics.
Users are classified into high (composite $\geq$ 75th percentile), medium, and low ($\leq$ 25th percentile) drift groups.

\begin{figure}[t]
  \centering
  \includegraphics[width=0.9\columnwidth]{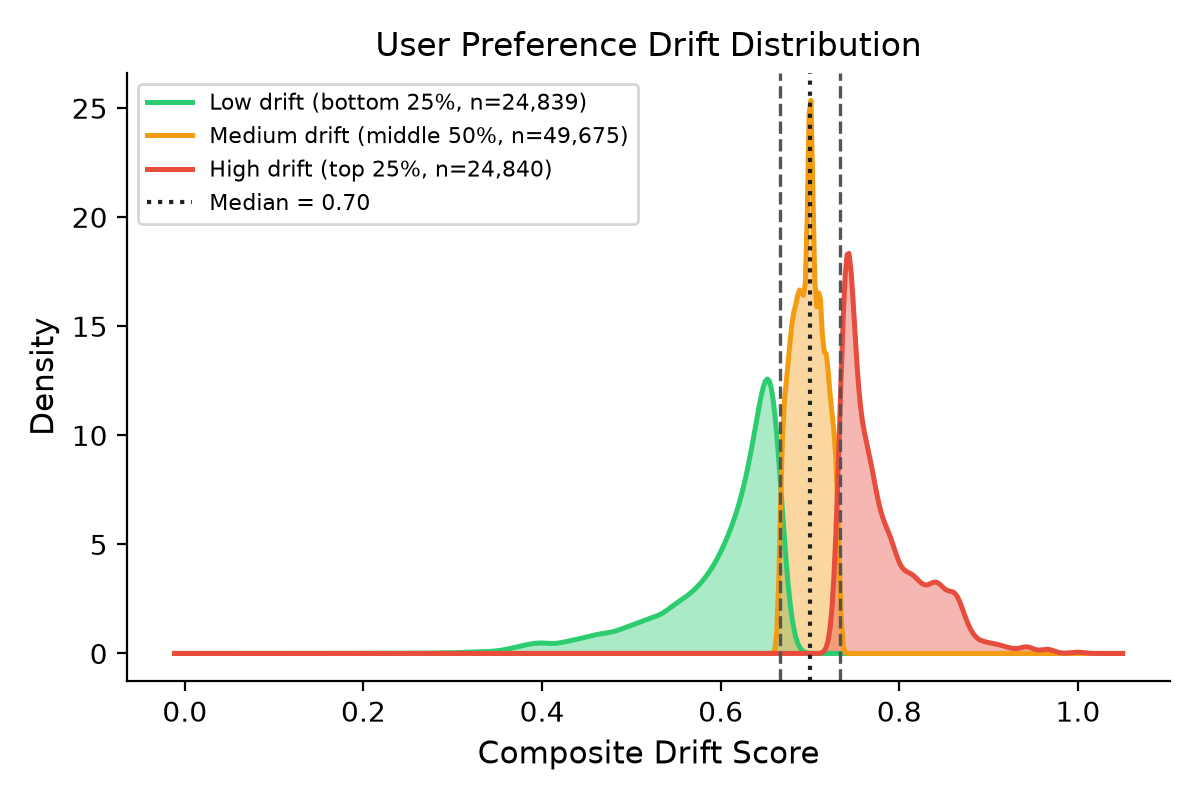}
  \caption{Distribution of composite preference drift scores across 99,354 Amazon users.
    Groups are defined by the 25th (low) and 75th (high) percentile thresholds (dashed lines).
    The median score of 0.70 and the absence of users near zero confirm that preference
    evolution is pervasive across all user segments.}
  \label{fig:drift_summary}
\end{figure}

On the Electronics and Clothing\_Shoes\_and\_Jewelry categories of Amazon Reviews 2023, the mean category JSD is 0.795 and the median novelty rate is 1.0 (i.e., the typical user interacts exclusively with new items in every window).
These figures confirm that preference evolution is not a rare edge case but a dominant behavioral pattern, motivating evolution-aware modeling for all users (Figure~\ref{fig:drift_summary}).

Table~\ref{tab:drift_eval} reports test-set HR@10, NDCG@10, HR@20, and MRR stratified by drift group, comparing MEMOIR against SASRec (strongest ID-based baseline), DuoRec, and UniSRec (strongest overall baseline, per Table~\ref{tab:main_results}).

\begin{table}[t]
\caption{Performance stratified by preference drift group (Amazon Reviews). Best per column within each group in \textbf{bold}, second best \underline{underlined}.}
\label{tab:drift_eval}
\small
\begin{tabular}{llcccc}
\toprule
\textbf{Drift} & \textbf{Model} & \textbf{HR@10} & \textbf{NDCG@10} & \textbf{HR@20} & \textbf{MRR} \\
\midrule
\multirow{4}{*}{High}
  & SASRec  & 0.1017 & \underline{0.0623} & 0.1302 & 0.0559 \\
  & DuoRec  & 0.0605 & 0.0397 & 0.0803 & 0.0396 \\
  & UniSRec & \textbf{0.1158} & 0.0612 & \textbf{0.1789} & \underline{0.0570} \\
  & MEMOIR  & \underline{0.1078} & \textbf{0.0653} & \underline{0.1554} & \textbf{0.0627} \\
\midrule
\multirow{4}{*}{Medium}
  & SASRec  & 0.0918 & 0.0565 & 0.1229 & 0.0519 \\
  & DuoRec  & 0.0550 & 0.0354 & 0.0740 & 0.0358 \\
  & UniSRec & \textbf{0.1099} & \textbf{0.0610} & \textbf{0.1679} & \textbf{0.0581} \\
  & MEMOIR  & \underline{0.0967} & \underline{0.0578} & \underline{0.1427} & \underline{0.0563} \\
\midrule
\multirow{4}{*}{Low}
  & SASRec  & 0.0800 & 0.0478 & 0.1121 & 0.0446 \\
  & DuoRec  & 0.0486 & 0.0305 & 0.0674 & 0.0312 \\
  & UniSRec & \textbf{0.1286} & \underline{0.0703} & \textbf{0.1940} & \underline{0.0657} \\
  & MEMOIR  & \underline{0.1239} & \textbf{0.0714} & \underline{0.1861} & \textbf{0.0673} \\
\bottomrule
\end{tabular}
\end{table}

A consistent pattern emerges, and it mirrors the aggregate split in Table~\ref{tab:main_results} rather than contradicting it: UniSRec leads the volume-oriented HR@10/HR@20 metrics in \emph{every} drift stratum, including both extremes.
MEMOIR's specific edge is on ranking-quality metrics---NDCG@10 and MRR---and only at the high- and low-drift extremes; in the medium-drift band, which holds half the user population, UniSRec leads on all four metrics and MEMOIR's aggregate near-tie (Table~\ref{tab:main_results}) is fully attributable to this stratum's majority weight working against it elsewhere.
We do not yet have a mechanistic explanation for why MEMOIR's ranking-quality advantage concentrates at the drift extremes---the ablation study (Table~\ref{tab:ablation}) shows this is not attributable to any single tested component in aggregate, and we have not re-run that ablation stratified by drift group, which would be a natural next step.
One plausible connection is the directional-consistency assumption discussed in §\ref{sec:limitations}: it is best suited to users with a clear, monotonic drift direction (arguably true of both high-drift users mid-transition and low-drift users with a stable direction), and may be neutral-to-harmful for the oscillating, less-directional preference changes more typical of the medium-drift band---but we surface this as a hypothesis for future ablation, not a demonstrated mechanism.

\section{Conclusion}

We presented MEMOIR, a framework that segments user history into temporal windows of LLM-generated behavioral memory and aggregates current state, evolution direction, and predicted future into a user representation.
On Amazon Reviews, MEMOIR clearly outperforms six traditional and LLM-enhanced baselines, and is statistically tied with the strongest baseline, UniSRec, in aggregate NDCG@10.
Neither result decomposes cleanly: an ablation over MEMOIR's own components shows none individually explains the gain over SASRec (§\ref{sec:ablation}), and stratifying by preference drift shows the near-tie with UniSRec masks a real, specific pattern---MEMOIR leads on ranking-quality metrics at the high- and low-drift extremes, while UniSRec leads throughout the medium-drift majority and on volume-oriented metrics everywhere (§\ref{sec:drift_analysis}).
We report the drift-stratified result, not the aggregate comparison or any single architectural ablation, as this paper's most substantive finding, and leave open why MEMOIR's advantage concentrates where it does.
Known limitations include the directional-consistency assumption that may penalize users with oscillating seasonal preferences---plausibly connected to the medium-drift band's weaker results, though not yet tested directly---and reduced evolution signal for sparse users with fewer than three windows (§\ref{sec:limitations}).
More broadly, MEMOIR trades the inference simplicity and model-agnostic flexibility of ID-based approaches---where lightweight backbones such as SASRec or GRU4Rec serve recommendations in single-digit milliseconds and contrastive or LLM-based augmentation can be applied as a plug-in training signal---for richer evolution-aware representations that require LLM-derived window embeddings.
In latency-critical production settings where sub-millisecond serving is required, or when the deployment stack must remain backbone-agnostic, these ID-based methods with semantic augmentation remain a practical and competitive alternative.
Future work includes adaptive window granularity to handle irregular activity patterns, scaling to multiple domains, and exploring inference-time efficiency through incremental memory caching to narrow the latency gap with ID-based pipelines.

\bibliographystyle{ACM-Reference-Format}
\bibliography{references}

\end{document}